\documentclass[preprint,showpacs,preprintnumbers,superscriptaddress,amsmath,amssymb]{revtex4}

\usepackage{graphicx}
\usepackage{dcolumn}
\usepackage{bm}

\usepackage{amssymb,bm,color}

\def\eqref#1{Eq.~(\ref{#1})}

\def\Eq#1{\begin{equation} #1 \end{equation}}
\def\Eqr#1{\begin{eqnarray} #1 \end{eqnarray}}
\def\Eqrsubl#1#2{\begin{subequations}\label{#1}\Eqr{#2}\end{subequations}}

\newcommand{\nn}{\nonumber}
\newcommand{\pd}{\partial}

\def\Xsp{{\rm X}}

\def\Ysp{{\rm Y}}

\def\X5sp{{\rm X}_5}
\def\Y3sp{{\rm Y}_3}
\def\Z3sp{{\rm Z}_3}

\def\lap{{\triangle}}
\def\e{{\rm e}}
\newcommand{\vect}[1]{\!\!\!\mbox{ \boldmath $#1$}}

\begin{document}

\preprint{KU-TP 045}

\title{Dynamical $p$-branes with a cosmological constant}

\author{Kei-ichi Maeda}%
\email{maeda"at"waseda.jp}
\affiliation{%
Department of Physics and RISE, Waseda University, Okubo 3-4-1,
Shinjuku, Tokyo 169-8555, Japan}%

\author{Masato Minamitsuji}%
\email{masato.minamitsuji"at"kwansei.ac.jp}
\affiliation{
Department of Physics, Graduate School of Science and Technology,\\
Kwansei Gakuin University, Sanda 669-1337, Japan
}%

\author{Nobuyoshi Ohta}
 \email{ohtan"at"phys.kindai.ac.jp}

\author{Kunihito Uzawa}
\affiliation{%
Department of Physics, Kinki University,
Higashi-Osaka, Osaka 577-8502, Japan
}%

\date{\today}

\begin{abstract}

We present a class of dynamical solutions in a
$D$-dimensional gravitational theory coupled to a dilaton,
a form field strength, and a cosmological constant.
We find that for any $D$ due to the presence of a cosmological constant, 
the metric of solutions depends on a quadratic function of the brane
world volume coordinates, and
the transverse space cannot be Ricci flat except for the case of 1-branes.
We then discuss the dynamics of 1-branes in a $D$-dimensional spacetime.
For a positive cosmological constant, 1-brane solutions with $D>4$
approach the Milne universe in the far-brane region.
On the other hand, for a negative cosmological constant,
each 1-brane approaches the others as the time evolves from a positive value,
but no brane collision occurs for $D>4$, since the spacetime close to
the 1-branes eventually splits into the separate domains.
In contrast, the $D=3$ case provides an example of colliding 1-branes.
Finally, we discuss the dynamics of 0-branes and show that for $D>2$, 
they behave like the Milne universe after the infinite cosmic time has passed.

\end{abstract}

\maketitle


\section{Introduction}
 \label{sec:introduction}

Supergravity is an important framework to study spacetime-dependent
solutions and their application to cosmology,
because it is a low-energy effective theory of superstrings.
Spacetime-dependent brane solutions in supergravity theories have played an
important role in the development of higher-dimensional
gravity theory~\cite{Lu:1996jk,Lu:1996er,Chen:2002yq,Ohta:2003uw,
Gibbons:2005rt,Kodama:2005fz,Kodama:2005cz,Binetruy:2007tu,Ishino:2005ru,Maeda:2009tq,
Maeda:2009zi,Maeda:2009ds,Gibbons:2009dr,Maeda:2010yk,Maeda:2010ja}.
Their central importance in supergravity theory can be anticipated
in their applications to cosmology and dynamical black holes.

The time-dependent generalizations of single static $p$-brane solutions
were discussed in \cite{Lu:1996jk,Lu:1996er,Chen:2002yq,Ohta:2003uw}.
Those with dependence on both time and space have been first discussed in
the case of a ten-dimensional type IIB supergravity model~\cite{Gibbons:2005rt}.
The extension of static solutions~\cite{Ohta} to the spacetime-dependent case
for $p$-brane and intersecting branes in the ten- or eleven-dimensional
supergravity are now well understood~\cite{Binetruy:2007tu,Ishino:2005ru,
Maeda:2009tq,Maeda:2009zi}.
In particular, the explicit form of the warp factor in the metric
has been obtained.
The solutions are specified by the choice of the scalar and
gauge fields and the values of the exponent in the warp factor of the metric.
The solutions give the Friedmann-Lema\^itre-Robertson-Walker (FLRW)
universe when we regard the homogeneous and
isotropic part of the brane world volume as our four-dimensional spacetime,
whereas they provide black hole solutions in a FLRW universe when we
regard the bulk transverse space as our three-dimensional space
~\cite{Maeda:2009zi,Maeda:2009ds}.
It was found that the warp factor in the metric
is a linear function of the time except for the trivial or
vanishing dilaton, and then
even for the fastest expanding case, the power is too small
to give a realistic expansion law as in the matter dominated era or
in the radiation dominated era~\cite{Binetruy:2007tu, Maeda:2009zi}.
In order to find a realistic cosmic expansion,
we have to include additional matter fields.
Note that no cosmological constant is considered in these solutions.

In another line of development, the dynamical solutions
in six-dimensional Nishino-Salam-Sezgin (NSS) supergravity theory
\cite{Nishino:1984gk, Salam:1984cj, Nishino:1986dc,
Gibbons:2003di,Aghababaie:2003ar} with a positive cosmological constant
have been investigated
in~\cite{Maeda:1984gq,Maeda:1985es,tbdh1,tbdh2,Minamitsuji:2010fp},
including applications to brane world models.
An application of a static solution in the six-dimensional Romans
supergravity \cite{romans1,romans2}
(with a negative cosmological constant)
to brane world model was also discussed in \cite{Aghababaie:2003ar}.
These solutions are considerably different from the above class of
solutions because of the presence of the cosmological constant.
Some arguments on the origin of the cosmological constant in the context
of string theory are given in \cite{AGMV,POL}.
These dynamical solutions cannot be in general derived from the ordinary ansatz
of fields used in a $p$-brane system.
One particular construction of dynamical solutions
was discussed recently in the NSS model and then
applied to brane world models in \cite{Minamitsuji:2010fp}.
In the present paper, dynamical 1-brane solutions in the NSS model
and in a class of the six-dimensional Romans supergravity
will be derived and used to understand the brane collisions.
Brane collisions in the special case of $p$-branes
have been originally discussed in \cite{Gibbons:2005rt}.

The main purpose of the present paper is to unify these two
lines of development, by showing that the methods
that have already been used in analyzing the dynamical
$p$-brane system without the cosmological constant lead naturally
to the extension of $p$-branes in theories with the cosmological constant.
We show that quadratic functions of the time, more precisely of the world volume
coordinates, appear in the solutions as the warp factor due to the contribution
of a cosmological constant.
These are similar to the dynamical solutions without a cosmological constant
with a trivial or vanishing dilaton found in \cite{Binetruy:2007tu}.
We apply the resulting dynamical solutions to brane collision and cosmology.
We also find that the dynamical 0-brane solution describes the Milne universe.

This paper is organized as follows.
In Sec.~\ref{sec:Dp}, we argue that there exists a procedure allowing one
to construct dynamical $p$-brane solutions with a cosmological constant
in a $D$-dimensional theory which generalizes the approach in \cite{Minamitsuji:2010fp},
and discuss their applications.
First in Sec.~\ref{sec:base}, we introduce our theory and derive
the basic equations, and reduce them to a set of simple equations
which should be satisfied for dynamical $p$-brane solutions.
They are given without specifying the brane world volume and other part of
the spacetime. We find that the basic difference of the solutions is that they
involve Einstein space for a nonvanishing cosmological constant, except for
$p=1$ and $p=0$.
Secondly,
in Sec.~\ref{D1}, we discuss the behavior of multiple 1-branes in our
broad class of solutions in the $D$-dimensional theory
and show that the solutions have interesting behaviors.
For a positive cosmological constant and $D>4$, 1-brane solutions
approach the Milne universe in the far region.
For a negative cosmological constant,
the spacetime starts with the structure of combined 1-branes, 
but a part of it eventually splits into separate regions as the time
increases from zero for $D>4$, which is similar to the result
in~\cite{Gibbons:2005rt}. Thus, 1-branes never collide.
In contrast, the case of $D=3$ provides an example of colliding branes.
Finally, in Sec.~\ref{sec:D0}, we discuss the
dynamical 0-brane solutions in the context of cosmology
and show that they give the Milne universe in $D>2$ dimensions.
Section \ref{sec:Conclusion} is devoted to our conclusions.


\section{Dynamical $p$-brane solutions with a cosmological constant}
\label{sec:Dp}

In this section, we consider dynamical $p$-brane solutions in theories
with the cosmological constant in $D$ dimensions.
First, we write down the equations of motion under a particular
ansatz for the metric, which is a generalization of the
known static $p$-brane solutions.
Then, we solve the equations of motion
and present the solutions explicitly for the cases of $p=1$ and 0.

\subsection{Basic equations and ``general" solutions}
\label{sec:base}

We consider a gravitational theory with the metric $g_{MN}$,
the dilaton $\phi$, the cosmological constant $\Lambda$,
and the antisymmetric tensor field of rank $(p+2)$, $F_{(p+2)}$.
The action which we consider is given by
\Eq{
S=\frac{1}{2\kappa^2}\int \left[\left(R-2\e^{\alpha\phi}
\Lambda\right)\ast{\bf 1}_D
 -\frac{1}{2}d\phi \wedge \ast d\phi
 -\frac{1}{2\cdot (p+2)!}\e^{\epsilon c\phi}
 F_{(p+2)}\wedge\ast F_{(p+2)}\right],
\label{D:action:Eq}
}
in the Einstein frame
where $\kappa^2$ is the $D$-dimensional gravitational constant,
$\ast$ is the Hodge operator in the $D$-dimensional spacetime,
$F_{(p+2)}$ is the $(p+2)$-form field strength,
and $c$, $\epsilon$, $\alpha$ are constants given by
\Eqrsubl{D:parameters:Eq}{
c^2&=&4-\frac{2(p+1)(D-p-3)}{D-2},
   \label{D:c:Eq}\\
\epsilon&=&\left\{
\begin{array}{cc}
 +&~{\rm if}~~p-{\rm brane~is~electric}\\
 -&~~~{\rm if}~~p-{\rm brane~is~magnetic}
\end{array} \right.
 \label{D:epsilon:Eq}\\
\alpha&=&-2\left(\frac{p+1}{D-2}\right)\left(\epsilon c\right)^{-1}.
    \label{D:alpha:Eq}
   }
The field strength $F_{(p+2)}$ is given by the
$(p+1)$-form gauge potential $A_{(p+1)}$
\Eq{F_{(p+2)}=dA_{(p+1)}.}

The actions of supergravities in $D=11$ and $D=10$ correspond to
$\Lambda=0$ case in (\ref{D:action:Eq}).
The bosonic part of the action of $D=11$ supergravity includes only 4-form
 ($p=2$)
without the dilaton. For $D=10$, the constant $c$ is precisely the dilaton
coupling for the Ramond-Ramond $(p+2)$-form in the type II supergravities.
Moreover, the action (\ref{D:action:Eq}) without the cosmological constant
also represents the leading-order expression for the low-energy limit of the
$D$-dimensional bosonic string. The bosonic string suffers from a
conformal anomaly unless $D=26$, which generates an additional
term in the effective action ~\cite{POL, Callan:1985ia, Cvetic:2000dm}.
For $D\ne 26$, the $D$-dimensional action can be given by
(\ref{D:action:Eq}) with $p=1$.

On the other hand,
the action (\ref{D:action:Eq}) for $D=6$ is related to the
six-dimensional supergravity theory. The bosonic part of the six-dimensional
action for NSS theory~\cite{Nishino:1984gk, Salam:1984cj, Nishino:1986dc}
and Romans theory \cite{romans1} are given
by the expression (\ref{D:action:Eq}) with positive 
and negative cosmological constants, respectively.

After varying the action with respect to the metric, the dilaton,
and the $(p+1)$-form gauge field, we obtain the field equations
\Eqrsubl{D:field equations:Eq}{
&&\hspace{-1cm}R_{MN}=\frac{2}{D-2}\e^{\alpha\phi}\Lambda g_{MN}
+\frac{1}{2}\pd_M\phi \pd_N \phi\nn\\
&&\hspace{-1cm}~~~~+\frac{1}{2\cdot (p+2)!}\e^{\epsilon c\phi}
\left[(p+2)F_{MA_2\cdots A_{p+2}} {F_N}^{A_2\cdots A_{p+2}}
-\frac{p+1}{D-2}g_{MN} F^2_{(p+2)}\right],
   \label{D:Einstein:Eq}\\
&&\hspace{-1cm}d\ast d\phi-\frac{\epsilon c}{2\cdot (p+2)!}
\e^{\epsilon c\phi}F_{(p+2)}\wedge\ast  F_{(p+2)}
-2\alpha \e^{\alpha\phi}\Lambda\ast{\bf 1}_D=0,
   \label{D:scalar:Eq}\\
&&\hspace{-1cm}d\left[\e^{\epsilon c\phi}\ast F_{(p+2)}\right]=0.
   \label{D:gauge:Eq}
}

To solve the field equations, we assume that the $D$-dimensional metric
takes the form
\Eq{
ds^2=h^a(x, y)q_{\mu\nu}(\Xsp)dx^{\mu}dx^{\nu}
  +h^b(x, y)u_{ij}(\Ysp)dy^idy^j,
 \label{D:metric:Eq}
}
where $q_{\mu\nu}(\Xsp)$ is a $(p+1)$-dimensional metric which
depends only on the $(p+1)$-dimensional coordinates $x^{\mu}$,
and $u_{ij}(\Ysp)$ is the $(D-p-1)$-dimensional metric which
depends only on the $(D-p-1)$-dimensional coordinates $y^i$.
Here, the $\Xsp$ space represents the world volume directions,
while the $\Ysp$ space does the transverse space of the $p$-brane.
The parameters $a$ and $b$ are given by
\Eq{
a=-\frac{D-p-3}{D-2},~~~~b=\frac{p+1}{D-2}.
 \label{D:paremeter:Eq}
}
The metric form (\ref{D:metric:Eq}) is
a straightforward generalization of the case of a static $p$-brane
system with a dilaton coupling \cite{Lu:1995cs, Binetruy:2007tu}.
Furthermore, we assume that the scalar field $\phi$ and
the gauge field strength $F_{(p+2)}$ are given by
\Eqrsubl{D:ansatz for fields:Eq}{
&&\e^{\phi}=h^{\epsilon c/2},
  \label{D:ansatz for scalar:Eq}\\
&&F_{(p+2)}=d(h^{-1})\wedge\Omega(\Xsp),
  \label{D:ansatz for gauge:Eq}
}
where $\Omega(\Xsp)$ denotes the volume $(p+1)$-form,
\Eq{
\Omega(\Xsp)=\sqrt{-q}\,dx^0\wedge dx^1\wedge \cdots \wedge
dx^p.
}
Here, $q$ is the determinant of the metric $q_{\mu\nu}$.

Let us first consider the Einstein Eqs.~(\ref{D:Einstein:Eq}).
Using the assumptions (\ref{D:metric:Eq}) and (\ref{D:ansatz for fields:Eq}),
the Einstein equations are given by
\Eqrsubl{D:cEinstein:Eq}{
&&\hspace{-1.5cm}R_{\mu\nu}(\Xsp)-h^{-1}\left(D_{\mu}D_{\nu} h
+\frac{2}{D-2}\Lambda q_{\mu\nu}\right)
-\frac{a}{2}h^{-1}
q_{\mu\nu}\left(\triangle_{\Xsp}h + h^{-1}\triangle_{\Ysp} h\right)=0,
 \label{D:cEinstein-mu:Eq}\\
&&\hspace{-1.5cm}h^{-1}\pd_{\mu}\pd_i h=0,
 \label{D:cEinstein-mi:Eq}\\
&&\hspace{-1.5cm}R_{ij}(\Ysp)-\frac{b}{2} u_{ij}\left(\triangle_{\Xsp} h
 +h^{-1}\triangle_{\Ysp}h \right)
 -\frac{2}{D-2}\Lambda u_{ij}=0
 \label{D:cEinstein-ij:Eq},
}
where $D_{\mu}$ is the covariant derivative with respect to
the metric $q_{\mu\nu}$,
$\triangle_{\Xsp}$ and $\triangle_{\Ysp}$ are
the Laplace operators on the space of
${\rm \Xsp}$ and the space ${\rm \Ysp}$, and
$R_{\mu\nu}(\Xsp)$ and $R_{ij}(\Ysp)$ are the Ricci tensors
of the metrics $q_{\mu\nu}$ and $u_{ij}$, respectively.
From Eq.~(\ref{D:cEinstein-mi:Eq}), we see that
the function $h$ must be in the form
\Eq{
h(x, y)= h_0(x)+h_1(y).
  \label{D:warp:Eq}
}
With this form of $h$, the other components of the Einstein Eqs.
(\ref{D:cEinstein-mu:Eq}) and (\ref{D:cEinstein-ij:Eq}) are rewritten as
\Eqrsubl{D:cEinstein2:Eq}{
&&R_{\mu\nu}(\Xsp)-h^{-1}\left(D_{\mu}D_{\nu} h_0
+\frac{2}{D-2}\Lambda q_{\mu\nu}\right)
-\frac{a}{2}h^{-1}
q_{\mu\nu} \left(\triangle_{\Xsp} h_0
+h^{-1}\triangle_{\Ysp}h_1\right)=0,
   \label{D:cEinstein-mu2:Eq}\\
&&R_{ij}(\Ysp)-\frac{b}{2}u_{ij}\left(\triangle_{\Xsp}h_0
+h^{-1}\triangle_{\Ysp}h_1\right)-\frac{2}{D-2}\Lambda u_{ij}=0.
   \label{D:cEinstein-ij2:Eq}
   }

Under the assumption (\ref{D:ansatz for gauge:Eq}),
the Bianchi identity is automatically satisfied.
The equation of motion for the gauge field~(\ref{D:gauge:Eq}) becomes
\Eqr{
d\left[\e^{\epsilon c\phi}\ast F_{(p+2)} \right]
&=&-\lap_{\Ysp}h_1\,\Omega(\Ysp)=0,
 }
where we have used (\ref{D:warp:Eq}), and $\Omega(\Ysp)$ is defined by
\Eq{
\Omega(\Ysp)=\sqrt{u}\,dy^1\wedge\cdots\wedge dy^{D-p-1}.
}
Hence, the gauge field equation gives
\Eq{
\lap_{\Ysp}h_1=0.
   \label{D:h1:Eq}
}

Let us next consider the scalar field equation.
Substituting Eqs.~(\ref{D:ansatz for fields:Eq}) and (\ref{D:warp:Eq}) into
Eq.~(\ref{D:scalar:Eq}), we obtain
\Eq{
\triangle_{\Xsp}h_0+8bc^{-2}\Lambda
+h^{-1}\triangle_{\Ysp}h_1=0.
  \label{D:scalar2:Eq}
}
Because of $\lap_{\Ysp}h_1=0$ in Eq.~(\ref{D:h1:Eq}), we are left with
\Eq{
\triangle_{\Xsp}h_0+8bc^{-2}\Lambda=0.
\label{D:solution for scalar:Eq}
}

Let us go back to the Einstein Eqs.~(\ref{D:cEinstein2:Eq}).
If $F_{(p+2)}=0$, the function $h_1$ becomes trivial.
On the other hand, for $F_{(p+2)}\ne 0$, the first term in Eq.~(\ref{D:cEinstein-mu2:Eq})
depends on only $x$ whereas the rest on both $x$ and $y$.
Thus Eqs.~(\ref{D:cEinstein2:Eq}) together with (\ref{D:h1:Eq}) and
(\ref{D:solution for scalar:Eq}) give
\Eqrsubl{D:Einstein2:Eq}{
&&R_{\mu\nu}(\Xsp)=0,~~~~
D_{\mu}D_{\nu}h_0+8\left[4(D-2)-2(p+1)(D-p-3)\right]^{-1}\Lambda q_{\mu\nu}=0,
   \label{D:Ricci-mn:Eq}\\
&&R_{ij}(\Ysp)+4(p-1)
\left[4(D-2)-2(p+1)(D-p-3)\right]^{-1}\Lambda u_{ij}=0.~~
   \label{D:Ricci-ij:Eq}
 }
If one solves these Eqs.~(\ref{D:Einstein2:Eq}) with Eq. (\ref{D:h1:Eq}),
the solution of the present system is given by Eqs. (\ref{D:metric:Eq})
and (\ref{D:ansatz for fields:Eq}) with (\ref{D:warp:Eq}).

Equation~(\ref{D:Ricci-mn:Eq}) implies that the function $h_0$ is the
same form as in the case of a single brane solution
with a trivial or vanishing dilaton as can be seen in \cite{Binetruy:2007tu}.
Thus, we find that the metric~(\ref{D:metric:Eq}) for the $p$-brane
in the system with
a cosmological constant is similar to that of the single D3-brane
or M-brane systems.
The difference from the $p$-brane metric consists in the $(D-p-1)$-dimensional
metric $u_{ij}$, which is affected by the existence of a cosmological constant:
For a nonvanishing cosmological constant, Eq.~(\ref{D:Ricci-ij:Eq})
implies that the $(D-p-1)$-dimensional space $\Ysp$ is an Einstein manifold.
The $(D-p-1)$-dimensional flat space is allowed only for $p=1$.
In the following two Secs.~\ref{D1} and \ref{sec:D0},
we shall discuss the $p=1$-brane solution in the flat $\Ysp$ space
and $p=0$-brane solutions in an Einstein space, respectively.

\subsection{The dynamical $1$-brane solution}
\label{D1}

Let us discuss the case of $p=1$, for which $\Ysp$ space is Ricci flat
 ($R_{ij}(\Ysp)=0$)
\cite{Minamitsuji:2010fp}.
We find that Eqs.~(\ref{D:Einstein2:Eq}) and (\ref{D:h1:Eq}) reduce to
\Eqrsubl{D1:Einstein:Eq}{
&&R_{\mu\nu}(\Xsp)=0,~~~~D_{\mu}D_{\nu}h_0+\Lambda q_{\mu\nu}=0,
   \label{D1:Ricci-mn:Eq}\\
&&R_{ij}(\Ysp)=0,~~~~\lap_{\Ysp}h_1=0.
   \label{D1:Ricci-ij:Eq}
 }
For the special case
\Eq{
q_{\mu\nu}=\eta_{\mu\nu}\,,
\quad u_{ij}=\delta_{ij}\,,
 \label{D:smetric:Eq}
 }
where $\eta_{\mu\nu}$ is the two-dimensional
Minkowski metric and $\delta_{ij}$ is
the $(D-2)$-dimensional flat Euclidean metric,
the solution for $h$ is obtained explicitly as
\Eq{
h(x, y)=-\frac{\Lambda}{2}x^{\mu}x_{\mu}+A_{\mu}x^{\mu}+B
+h_1(y)\,,
 \label{D:h2:Eq}
}
where $A_{\mu}$, $B$ are constant parameters and
the harmonic function $h_1$ is found to be
\Eqrsubl{harmonics:Eq}{
h_1(\,\vect{y})&=&\sum_{l=1}^{N}\frac{M_l}{|\,\vect{y}-\,\vect{y}_l|^{D-4}}
~~~~~~{\rm for}~~D\neq 4\,,
\label{harmonics_neq4}
\\
h_1(\,\vect{y})&=&\sum_{l=1}^{N}M_l\, \ln |\,\vect{y}-\,\vect{y}_l|
~~~~{\rm for}~~D= 4
\,.
\label{harmonics_4}
}
Here $|\vect{y}-\,\vect{y}_l|
=\sqrt{\left(y^1-y^1_l\right)^2+\left(y^2-y^2_l\right)^2+\cdots+
\left(y^{D-2}-y^{D-2}_l\right)^2}$, and
$M_l~ (l=1\cdots N)$ are mass constants of 1-branes
located at $~\vect{y}_l$.
The metric, dilaton, and gauge field strength of the solution are
 given by Eqs.~(\ref{D:metric:Eq}), (\ref{D:paremeter:Eq}) and
(\ref{D:ansatz for fields:Eq}), respectively.
In the case of $D=6$,
this solution for $\Lambda>0$ describes that in the NSS model
(see the Appendix of \cite{Minamitsuji:2010fp}),
while that for $\Lambda<0$ gives a dynamical $1$-brane solution
in the ${\cal N}=\tilde 4^{g}$ class
(following the classification in \cite{romans2}) of the
six-dimensional Romans supergravity \cite{romans1,romans2,Aghababaie:2003ar}
with vanishing $SU(2)$ and Abelian gauge field strengths.

We see that $D=4$ dimension is critical.
For $D\geq 4$, namely, if the number of transverse space is greater than two,
$h_1$ has an inverse power dependence on $|\vect{y}|$,
while for $D=3$ it is proportional to $|y|$.
This is because $h_1$ is the harmonic function on the
$(D-2)$-dimensional Euclid space Y,
which follows from the ansatz of the
metric~(\ref{D:metric:Eq}) and the form fields~(\ref{D:ansatz for fields:Eq}).
As we will discuss later, the difference in the transverse dimensions
causes significant difference in the behaviors of the gravitational field
strengths in the transverse space, and the possibility of brane collisions.

The metric obtained for the solution~(\ref{D:warp:Eq}) and (\ref{D:h2:Eq})
is not of the product-type. The origin of this property is the
existence of a nontrivial gauge field strength, which forces
the function $h$ to be a linear combination of a function of
$x^{\mu}$ and a function of $y^i$, unlike the conventional assumption.
The function (\ref{D:h2:Eq}) implies that we cannot drop the dependence
on the world volume coordinate for a nonvanishing cosmological constant.
This solution 
leads to the inhomogeneous universe owing to the function $h_1$
when we regard the bulk transverse space as our three-dimensional space.
If we consider the spacetime $\Xsp$ for $p\ge 3$ to contain
our four-dimensional universe, the scale factor of our Universe also
includes the inhomogeneity due to functions $h_0$ and $h_1$.
There are then two possibilities to obtain the four-dimensional homogeneous
and isotropic universe in the limit when the function $h_1$ is negligible.
One is the $p=0$ and $D=4$ case, which we discuss in Sec.~\ref{sec:D0}.
The other is the case that we live in the three-dimensional
transverse space Y after compactifying the $p$-dimensional
world volume. In this case, since we fix our Universe at
some position in the X space, the scale
factor of our Universe is proportional to the linear function of
the cosmic time in the four-dimensional Einstein frame,
giving the four-dimensional Milne universe. For $h_1\rightarrow 0$,
this is the same description as the ordinary Kaluza-Klein compactification.

Note that the $D$-dimensional metric (\ref{D:metric:Eq}) is regular for $h>0$
but has curvature singularities where $h=0$. So if it happens that $h$ changes
sign somewhere in the $D$-dimensional spacetime, a spacetime is restricted
to the $h>0$ region bounded by curvature singularities.
Around the $x=0$ plane, the spacetime appears to split into
disconnected regions, though it is not really separated in the whole.
We now show that this happens in our solutions.

The solution (\ref{D:h2:Eq}) with $N$ 1-branes takes the form
\Eqr{
ds^2&=&
\left[{\Lambda\over 2}
\left(t^2-x^2\right)+h_1(\,\vect{y})\right]^{-{D-4\over D-2}}
\left(-dt^2+dx^2\right)
+
\left[{\Lambda\over 2}
\left(t^2-x^2\right)+h_1(\,\vect{y})
\right]^{2\over D-2}d\,\vect{y}^2
\,,
   \label{D:sufrace0:Eq}
}
where we set $A_{\mu}=B=0$ in (\ref{D:h2:Eq}) and the function $h_1$ is
defined in (\ref{harmonics:Eq}).
The behavior of the harmonic function $h_1$ is classified into two
classes depending on the dimensions $D$, i.e.
{\it 1}. $D>4$, and  {\it 2}. $D=3$, which we will discuss
below separately.
For the remaining dimension $D=4$, the harmonic function $h_1$ diverges
both at infinity and near 1-branes.
In particular, because $h_1\rightarrow -\infty$,
there is no regular spacetime region near branes.
Hence, such solutions are not physically relevant.

\subsubsection{$D>4$}

First, we discuss the asymptotic structure.
Near branes, i.e., in the limit of $\,\vect{y}\rightarrow \,\vect{y}_l$,
the harmonic function $h_1$ becomes dominant.
Hence, we find a static 1-brane structure.
On the other hand, in the far-brane region, i.e.,
in the limit of $|\,\vect{y}|\rightarrow \infty$,
we find $h\approx {\Lambda\over 2}(t^2-x^2)$ because
$h_1$ vanishes. The metric is given by
\Eqr{
ds^2&=&
\left[{\Lambda\over 2}
\left(t^2-x^2\right)\right]^{-{D-4\over D-2}}
\left(-dt^2+dx^2\right)
+
\left[{\Lambda\over 2}
\left(t^2-x^2\right)
\right]^{2\over D-2}d\,\vect{y}^2
~~~~
\,,
   \label{D:sufrace:Eq}
}
which looks inhomogeneous at first glance.
However, if $\Lambda>0$, by the coordinate transformation
\begin{eqnarray}
&&t=\sqrt{2 \over \Lambda}\,
T \cosh X~~~,~~~~x=\sqrt{2 \over \Lambda}\,
T \sinh X
\,,
\end{eqnarray}
we find
\begin{eqnarray}
ds^2&=&{2\over \Lambda}\,T^{-{2(D-4)\over D-2}}
\left[-dT^2+T^2\left(dX^2+d\, \vect{Y}^2\right)
\right]
\nonumber \\
&=&{2\over \Lambda}\,\left[
-d\tau^2+\left({2\over D-2}\right)^2
\,\tau^2
\left(dX^2+d\,\vect{Y}^2\right)
\right]
\,,
\label{D:sufrace:Eq_p}
\end{eqnarray}
where
\begin{eqnarray}
\tau=\left({D-2\over 2}\right)T^{2\over D-2}\,,
\end{eqnarray}
and $\,\vect{Y}=\sqrt{\Lambda/2}~\vect{y}$.
Note that
${\Lambda\over 2}(t^2-x^2)=T^2$.

This metric (\ref{D:sufrace:Eq_p}) represents an isotropic and
homogeneous spacetime whose scale factor changes as the cosmic time
$\tau$, which is known as the Milne universe.
Hence, we can consider that the present solution
with $\Lambda>0$ describes a system of $N$
1-branes in the Milne universe.
The existence of the expanding Milne universe is guaranteed by
the scalar field with the exponential potential.
When we have a scalar field with an exponential potential
$V\propto \exp (\alpha\phi)$, we find the FLRW universe solution,
whose scale factor expands as $t^\lambda$, where $\lambda=2/[(D-2)\alpha^2]$
\cite{Halliwell:1987}.
Equation (\ref{D:alpha:Eq})
 gives the present coupling constant as $\alpha^2=2/(D-2)$, finding
$\lambda=1$, which corresponds to the Milne universe.

On the other hand, if $\Lambda<0$,
we should perform the following coordinate transformation:
\begin{eqnarray}
&&t=\sqrt{-{2 \over \Lambda}}\,
X \sinh T~~,~~~x=\sqrt{-{2 \over \Lambda}}\,
X \cosh T
\,.
\end{eqnarray}
We find
\begin{eqnarray}
ds^2&=&{2\over |\Lambda|}\,X^{-{2(D-4)\over D-2}}
\left[dX^2+X^2\left(-dT^2+d\, \vect{Y}^2\right)
\right]
\nonumber \\
&=&{2\over |\Lambda|}\,\left[
d\xi^2+\left({2\over D-2}\right)^2
\,\xi^2
\left(-dT^2+d\,\vect{Y}^2\right)
\right]
\,,
\label{D:sufrace:Eq_m}
\end{eqnarray}
where
\begin{eqnarray}
\xi=\left({D-2\over 2}\right)X^{2\over D-2},
\end{eqnarray}
and $\,\vect{Y}=\sqrt{|\Lambda|/2}~\vect{y}$.
Note that
${\Lambda\over 2}(t^2-x^2)=X^2$.
This metric (\ref{D:sufrace:Eq_m}) describes a conformally flat
and static inhomogeneous spacetime, which looks similar to a Milne universe
but is not a cosmological solution.

Next, we analyze a system of two 1-branes, which are located at
$\,\vect{y}=(\pm L, 0,\cdots,0)$.
Since the behavior of spacetime highly depends on the signature of
a cosmological constant, we discuss the dynamics separately.
\\

(1) $\Lambda>0$

As we have mentioned above, the metric function is singular at zeros
of the solution (\ref{D:h2:Eq}). Namely
the regular spacetime exists inside the domain restricted by
\Eq{
h(t,x, \bm y) \equiv h_1(\,\vect{y})+\frac{1}{2}\Lambda \left(t^2-x^2\right)>0,
}
where the function $h_1$ is defined in (\ref{harmonics:Eq}).
The spacetime cannot be extended beyond this region, because the scalar field
$\phi$ diverges, giving rise to a curvature singularity.

\begin{figure}[h]
\begin{center}
\includegraphics[scale=.8]{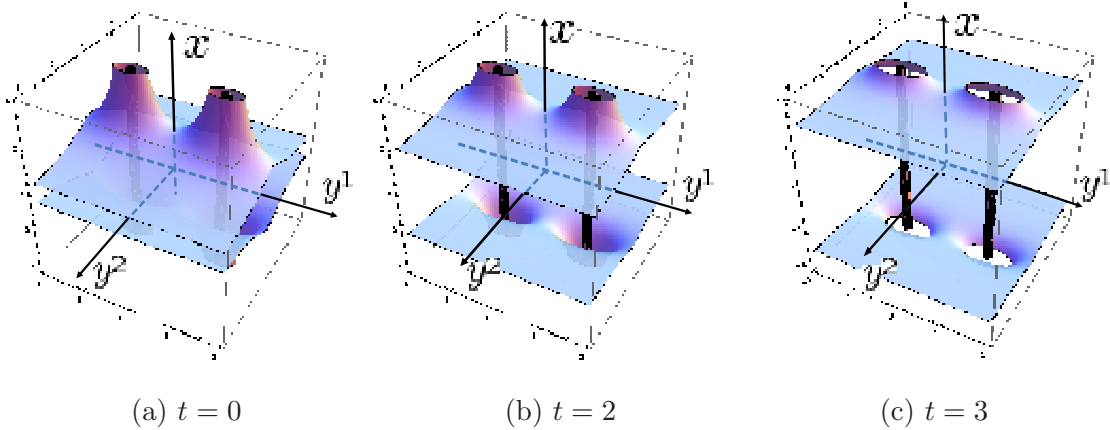}
\\
(a) $t=0$\hskip 3.5cm
(b) $t=2$\hskip 3.5cm
(c) $t=3$
\caption{\baselineskip 12pt
The time evolution of the singular hypersurfaces $h=0$
for two equal-mass 1-branes in $D=6$ and $\Lambda>0$.
Our parameters are $\Lambda=1$, $L=1$, and $M=1$.
We depict the surfaces with $y^3=y^4=0$.
The regular part of the spacetime corresponds to
that between these hypersurfaces.
}
\label{positive}
\end{center}
\end{figure}

\begin{figure}[h]
\begin{center}
\includegraphics[width=6cm,clip]{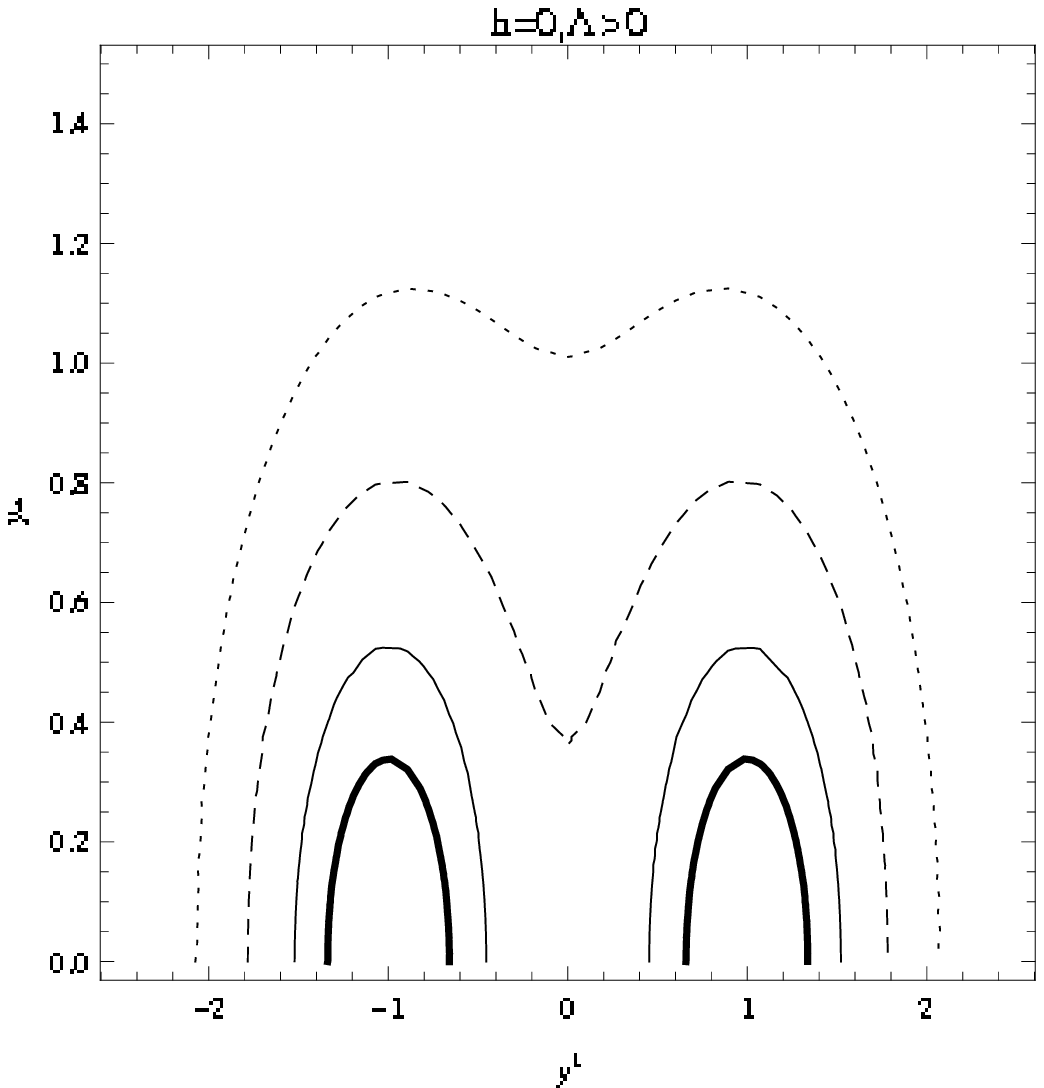}
\hskip 1cm
\includegraphics[width=6cm,clip]{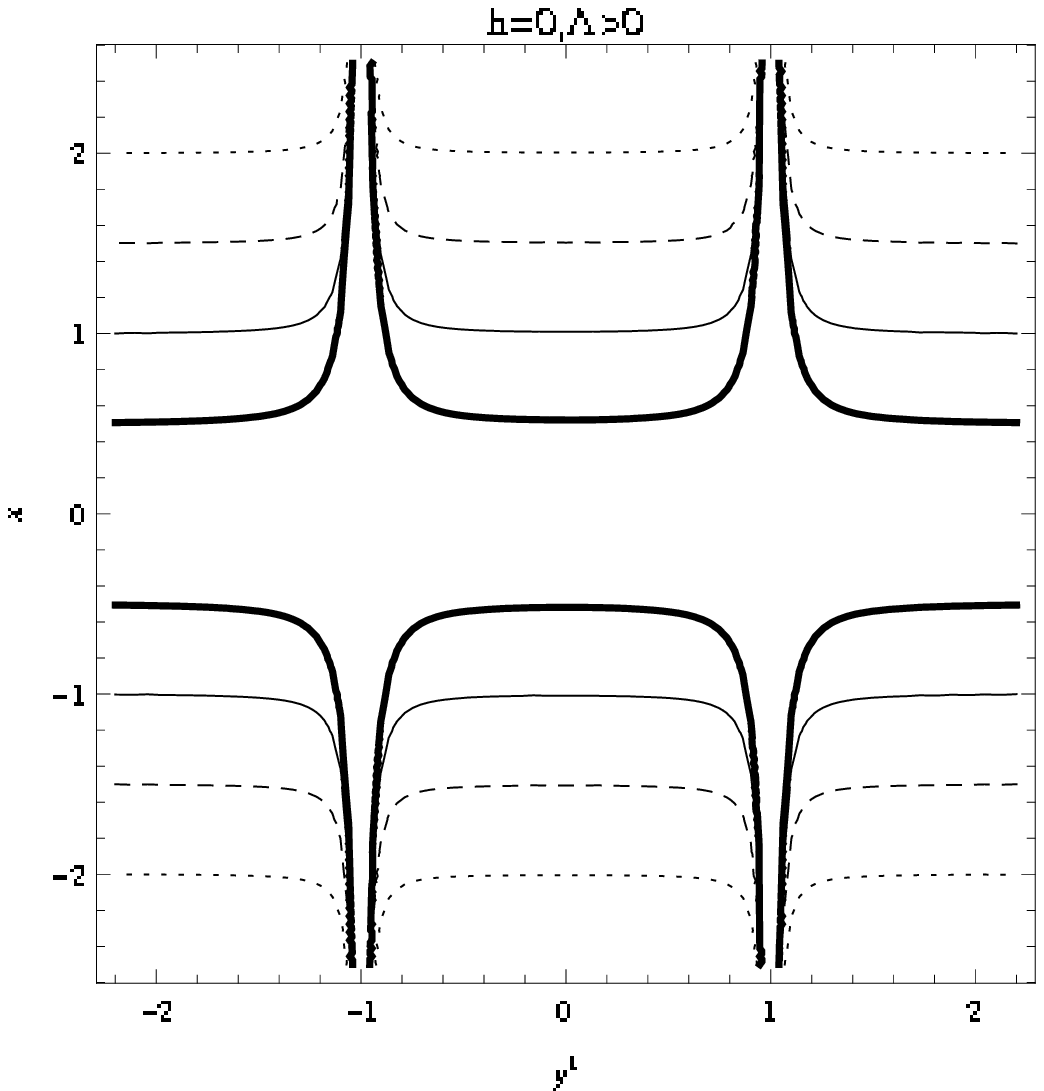}\\
(a) $x=0$-plane \hskip 4.3cm
(b) $y_\perp=0$-plane
\caption{\baselineskip 12pt
The contours show the level set $h=0$ at  $y_{\perp}=0$ for
two 1-branes in the case of $D=6$ and $\Lambda>0$.
The thick solid lines correspond to the surface of the
$h=0$ at the initial positive time. The lines change to solid,
dashed, and then dotted ones
as the time evolves.
As the time progresses, the domain expands.
Hence, the two level sets approach each other.
}
\label{positive2}
\end{center}
\end{figure}

In the case of $D=6$, we illustrate the positions of
two equal-mass 1-branes and time evolution of the
singular hypersurfaces $h=0$ in Fig.~\ref{positive}.
The regular part of the spacetime corresponds to that
between these hypersurfaces.
Two cross sections of the singular hypersurfaces ($x=0$ and $y_\perp=0$)
for $D=6$ are also shown in Fig. \ref{positive2},
where $y_{\perp}$ is defined by
\Eq{
y_{\perp}=\sqrt{\left(y^2\right)^2
+\cdots
+\left(y^{D-2}\right)^2}\,.
}
This case has the time-reversal symmetry.
Hence, 
the evolution for $t<0$ is obtained by the time-reversal transformation
[For $t<0$,
(c) $\rightarrow$ (b) $\rightarrow$ (a) in Fig. \ref{positive}].

The regular spacetime with two 1-branes ends on these singular
hypersurfaces.
First, we consider the period of $t>0$.
Initially ($t=0$), the regular region is small, but it increases in time.
The $x=0$ plane is always regular.
In the large-$|x|$ region,
two branes are disconnected by a singularity.
The metric (\ref{D:sufrace:Eq})
implies that the transverse dimensions ($y_\perp$) expand
 asymptotically
as $\tilde \tau$, while the
spatial dimension of the world volume ($x$) contracts
 asymptotically
as $\tilde \tau^{-(D-4)/2}$ for
 fixed spatial coordinates
($x$ and $\,\vect{y}$),
where $\tilde \tau$ is the proper time of the coordinate observer.
However, it is observer-dependent.
As we mentioned before, it is static near branes, and
the spacetime approaches a Milne
universe in the far region ($|\,\vect{y}|\rightarrow \infty$),
which expands in all directions isotropically.
For the period of $t<0$, the behavior of spacetime is
the time reversal of the period of $t>0$.

\begin{figure}[h]
\begin{center}
\includegraphics[scale=.4]{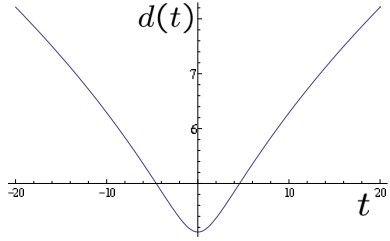}\\
\caption{\baselineskip 12pt
The proper distance between two branes at $x=0$
and $y_\perp=0$ for the case of $D=6$. We set $\Lambda=1$ and $M=1$.
Initially ($t<0$), the distance decreases, but turns to increase at $t=0$,
and then two 1-branes segregate each other.}
\label{fig:d_Lp}
\end{center}
\end{figure}

The proper distance at $x=0$ and $y_{\perp}=0$
between two branes is given by
\begin{eqnarray}
d(t)&=&\int_{-L}^L dy^1 \left[
{\Lambda\over 2}t^2+{M\over |y^1+L|^{D-4}}
+{M\over |y^1-L|^{D-4}}
\right]^{1\over D-2}
\nonumber \\
&=&
\left({ML^2}\right)^{1\over D-2}
\int_{-1}^1 d\eta
\left[
\left({\Lambda L^{D-4}\over 2M}\right)\,
t^2+{1\over |\eta+1|^{D-4}}
+{1\over |\eta-1|^{D-4}}\right]^{1\over D-2}
\,,
\label{distance}
\end{eqnarray}
which is a monotonically increasing function of $t^2$.
We show $d(t)$ integrated numerically in Fig. \ref{fig:d_Lp}
for the case of $D=6$.
It shows that two 1-branes are initially ($t<0$) approaching,
the distance $d$ takes the minimum finite value at $t=0$,
and then two 1-branes segregate each other.
They will never collide (see Fig. \ref{fig:d_Lp}).
\\

(2) $\Lambda<0$

\begin{figure}[h]
\begin{center}
\includegraphics[scale=.8]{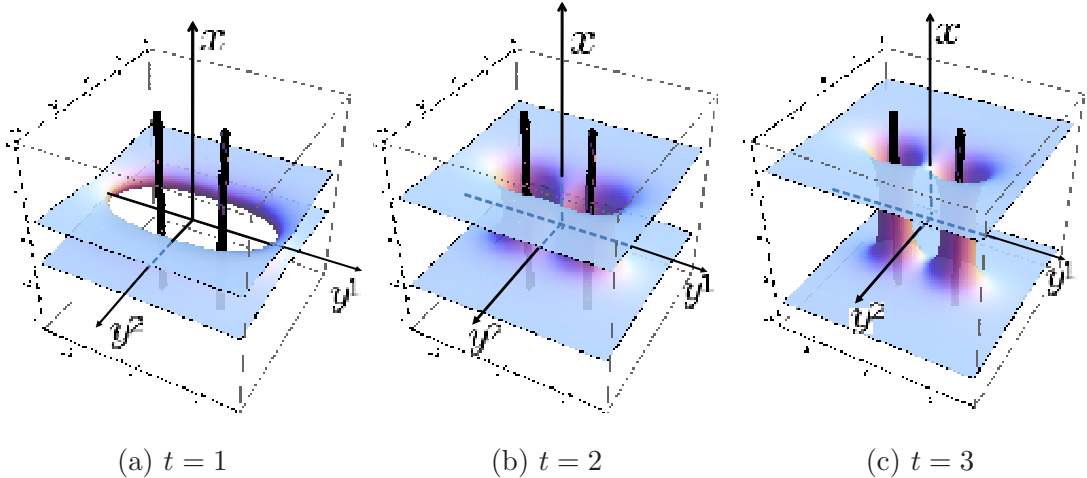}
\\
(a) $t=1$\hskip 3.5cm
(b) $t=2$\hskip 3.5cm
(c) $t=3$\\
\caption{\baselineskip 12pt
The position of two equal-mass 1-branes, and
the time evolution of the singular hypersurface $h=0$
with $y^3=y^4=0$ in the case of $D=6$ and $\Lambda<0$.
Our parameters are $\Lambda=-1$, $L=1$, and $M=1$.
The regular part of the spacetime is the region involving those
above and below the hypersurface.
}
\label{negative}
\end{center}
\end{figure}

\begin{figure}[h]
\begin{center}
\includegraphics[width=6cm,clip]{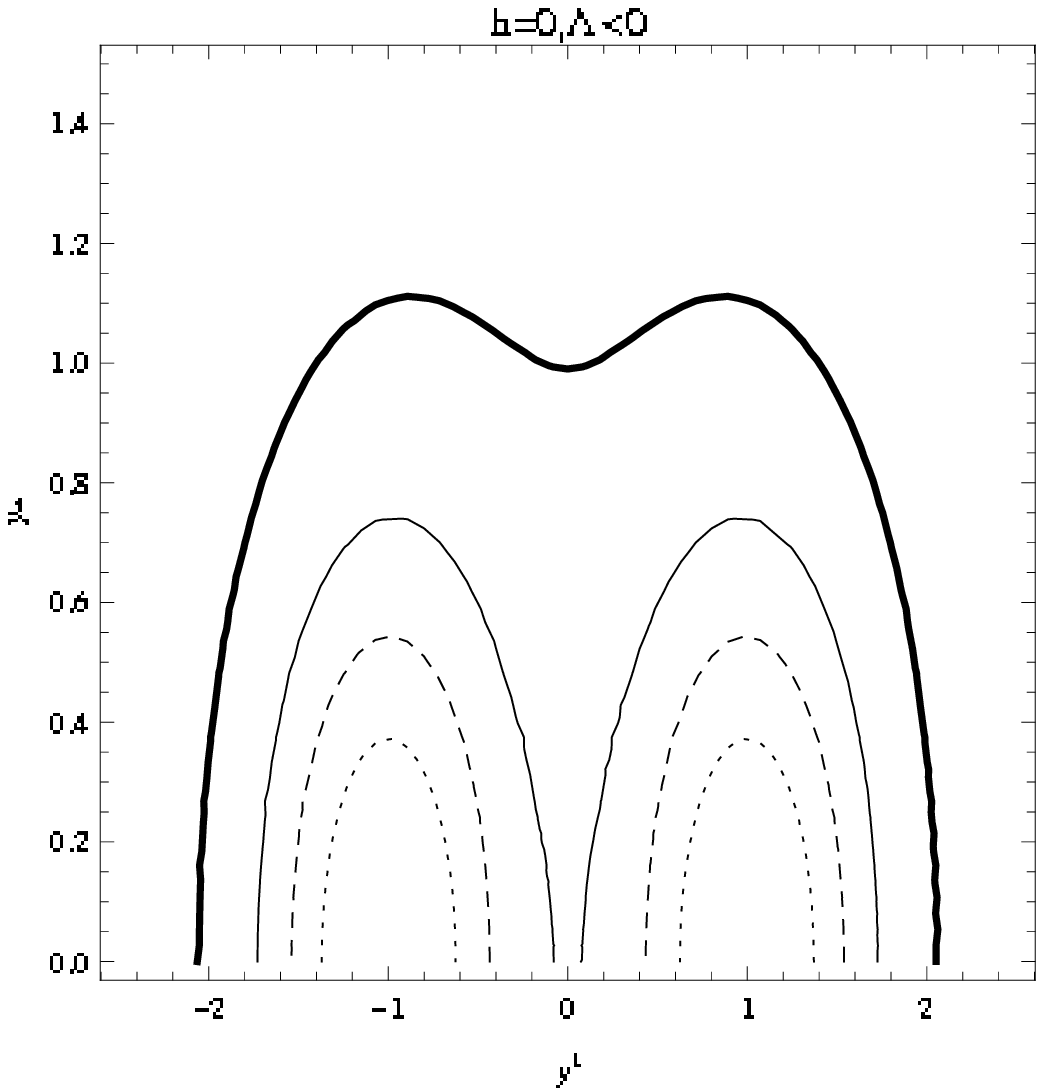}
\hskip 1cm
\includegraphics[width=6cm,clip]{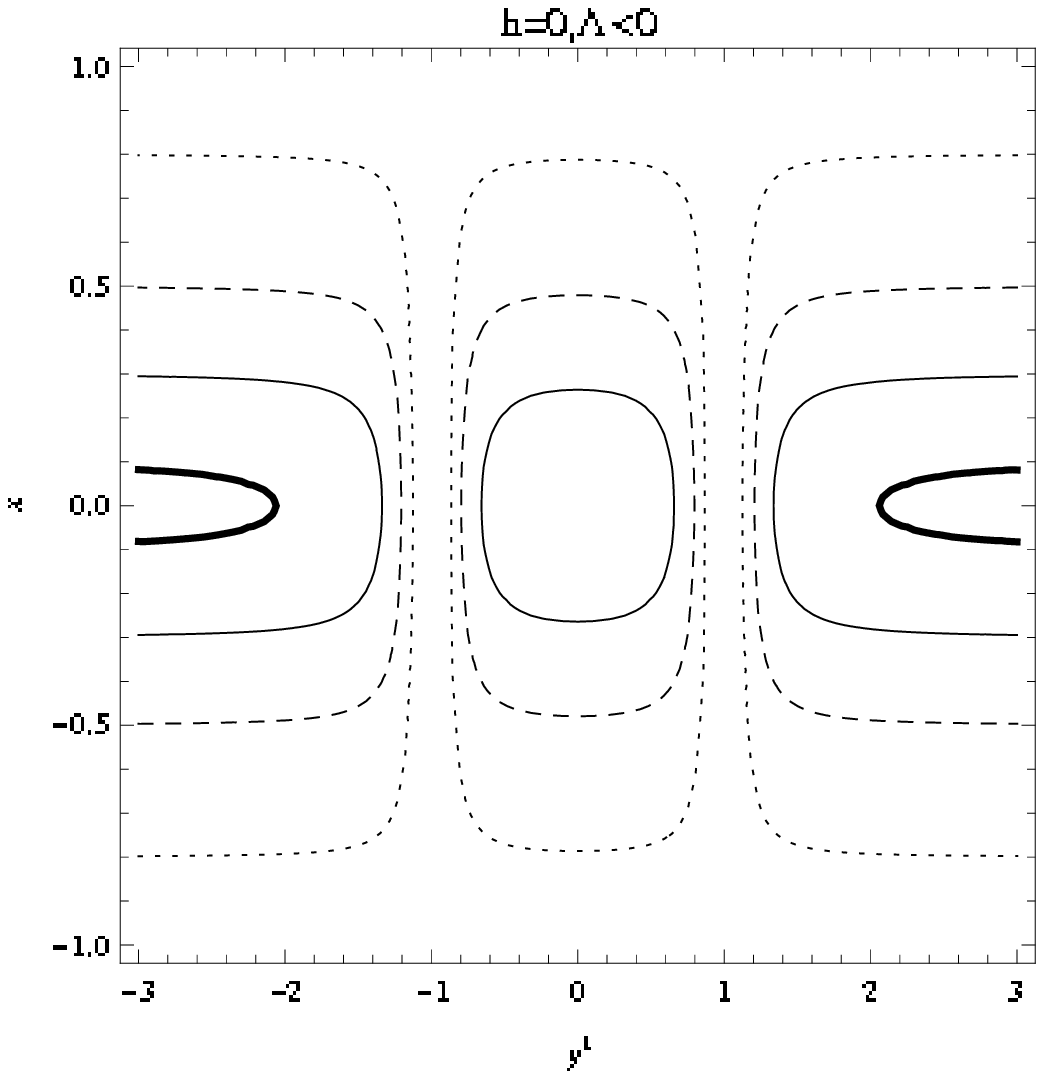}\\
(a) $x=0$-plane \hskip 4.3cm
(b) $y_\perp=0$-plane
\caption{\baselineskip 12pt
The contours show the level set $h=0$ at $x=0$ (a)
and $y_{\perp}=0$ (b) for
two 1-branes in the case of $D=6$ and $\Lambda<0$.
The thick solid lines correspond to the surface of the
$h=0$ at the initial positive time. The lines change to solid,
dashed, and then dotted ones
as the time evolves. The level set splits into two components,
which then shrink around the two 1-branes in time. }
\label{negative2}
\end{center}
\end{figure}

Next, we discuss the case of $\Lambda<0$.
We illustrate the positions of two equal-mass 1-branes and time evolution
of the singular hypersurface $h=0$ in Fig.~\ref{negative}
for the case of $D=6$.
The regular part of the spacetime is the region involving those
above and below the hypersurface.
Two cross sections of the singular hypersurface ($x=0$ and $y_\perp=0$)
are also shown in Fig.~\ref{negative2}.


First, we consider the period of $t\geq 0$.
Initially ($t=0$), 
all of the region of $(D-1)$-dimensional space is regular
except at $y\rightarrow\infty$ on the $x=0$ plane
[see Fig. \ref{negative}(a)].
As time evolves, the singular hypersurface erodes
the large $x$ region as shown in Fig. \ref{negative}.
The $\,\vect{y}$-coordinate region is also invaded in time.
As a result, only the region of
large-$x$  and near 1-branes remains regular.
When we watch this evolution on the $x=0$ plane,
the singular circle appears at infinity and
comes to the region of two branes [Fig.~\ref{negative}(a)].
A singular hypersurface eventually surrounds each 1-brane
individually and then the regular regions near 1-branes splits
into two isolated throats [see Figs.~\ref{negative} (b), (c)
 and \ref{negative2}].
For the period of $t<0$, we find the time-reversed
evolution of the case of $t>0$.

\begin{figure}[h]
\begin{center}
\includegraphics[scale=.4]{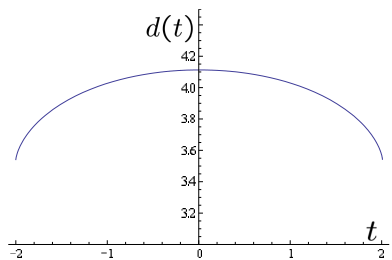}
\caption{The proper distance between two branes at $x=0$
and $y_\perp=0$ for the case of $D=6$. We set $\Lambda=-1$ and $M=1$.
Initially ($t<0$), the distance increases, but turns to decrease for $t>0$.
A singularity appears at $t=2$ when the distance is still finite.}
\label{fig:d_Lm}
\end{center}
\end{figure}
We also calculate the distance $d(t)$
at $x=0$ and $y_\perp=0$ between two branes before
the singularity appears.
The distance is also given by Eq. (\ref{distance}).
However in the present case, $d$ is a monotonically decreasing
function of $t^2$.
Hence, $d$ increases when $t<0$, but it turns to decrease after $t=0$.
We show the time change of the distance in Fig. \ref{fig:d_Lm}
for the case of $D=6$.
It could mimic brane collisions.
However, a singularity appears between two branes
before the distance vanishes, i.e.,
a singularity forms before collision of two branes.
Hence, we cannot discuss a brane collision in this example.
It is not the case if $D=3$, which we will discuss next.


\subsubsection{$D=3$}
Here we discuss the case of $D=3$, which may provide us
 a colliding 1-brane model. Although this is a toy model,
it may capture the essence of brane collision.
The dynamical 1-brane solution is written as
\Eqr{
ds^2&=&\left[\frac{1}{2}\Lambda\left(t^2-x^2\right)+
 \sum_{l}M_l|y-y_l|\right]\left(-dt^2+dx^2\right)\nn\\
 &&~~~+\left[\frac{1}{2}\Lambda\left(t^2-x^2\right)+
 \sum_{l}M_l|y-y_l|\right]^2dy^2,
   \label{3D:metric:Eq}
}
where the constant $y_l$ denotes the position of the 1-brane with
charge $M_l$.

Let us consider the two 1-branes with the brane charge $M_1$ at $y=0$
and the other $M_2$ at $y=L$.
The proper distance between the two 1-branes is given by
\Eqr{
d(t,x)&=&\int^L_0 dy \left[\frac{1}{2}\Lambda\left(t^2-x^2\right)+
 M_1|y|+M_2|y-L|\right]\nn\\
 &=&\frac{1}{2}L\left[\Lambda\left(t^2-x^2\right)+
 \left(M_1+M_2\right)L\right].
  \label{3D:length:Eq}
}
Equation~(\ref{3D:length:Eq}) implies that the branes collide
at $t_{\rm c}(x)=\pm\sqrt{-(M_1+M_2)L/\Lambda+x^2}$,
which depends on the brane charges and the place of the world volume.
On the collision, a singularity forms at $y=L/2$ due to $h=0$.

\begin{figure}[h]
\begin{center}
\includegraphics[width=6cm,clip]{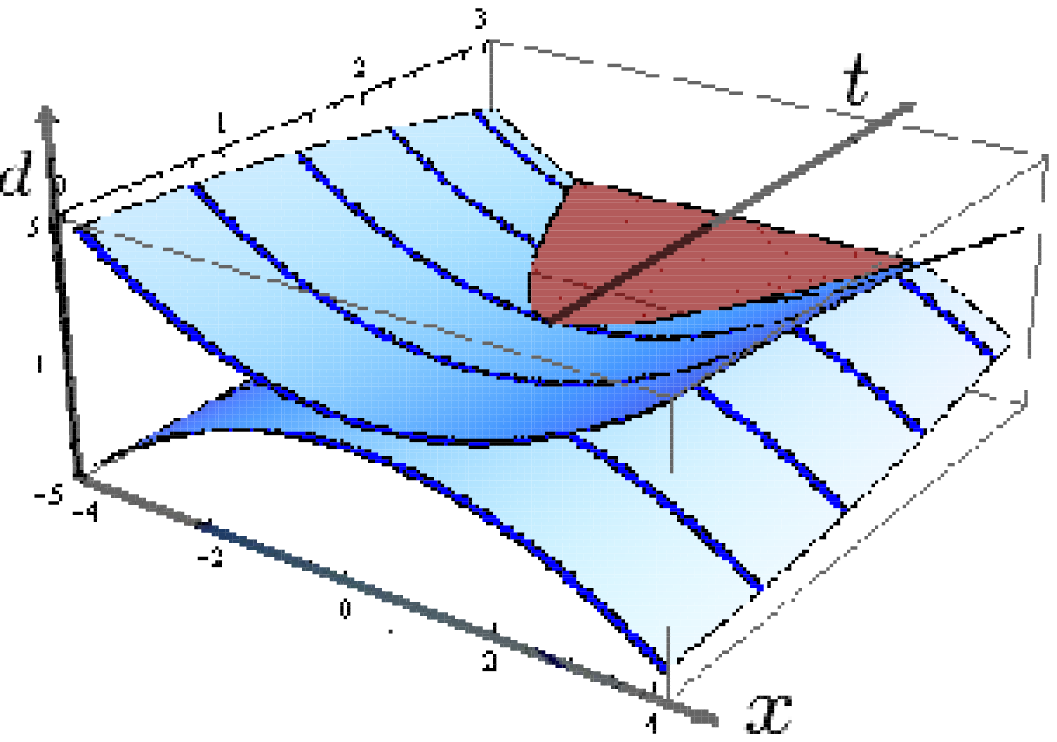}
\hskip 1cm
\includegraphics[width=6cm,clip]{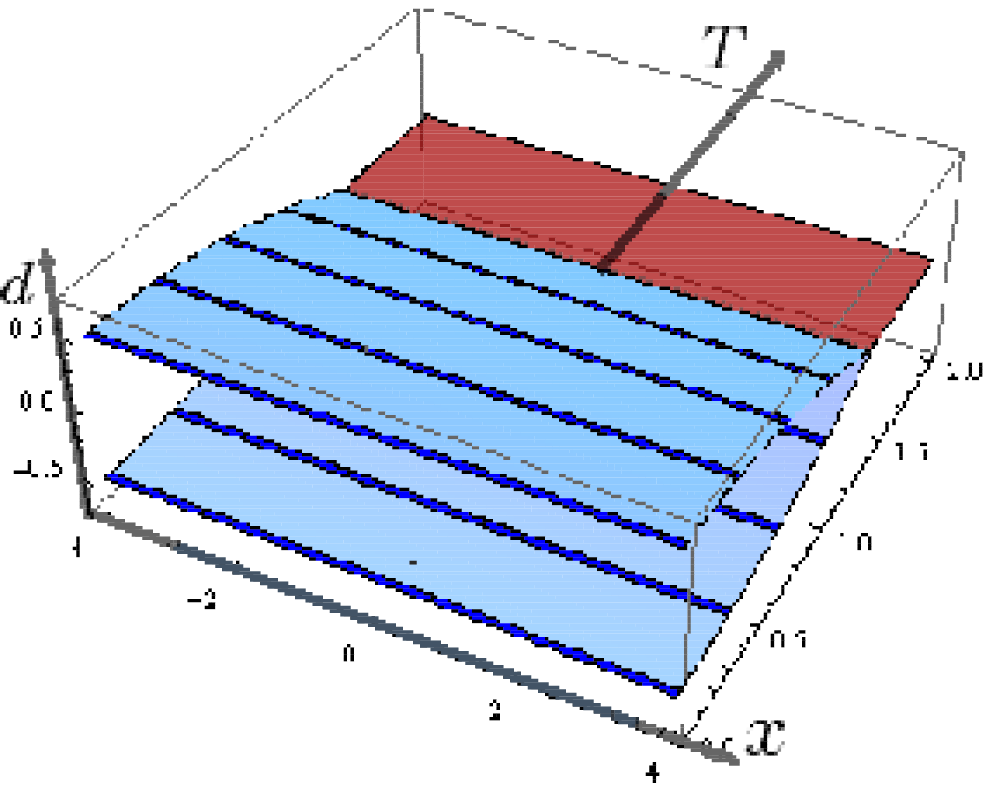}\\
(a) $t$-coordinate \hskip 4.3cm
(b) $T$-coordinate
\caption{\baselineskip 12pt
Collision of two 1-branes in three-dimensional spacetime with $\Lambda<0$.
(a) The world sheets of two 1-branes are shown
in the $(x, d, t)$ space, where
$d(t,x)$ is the proper distance between two 1-branes at $x$.
In order to see the motion of two branes,
the positions of 1-branes evolved with the equal
time interval are also shown by the blue (thick) curves.
The 1-branes collide first at the center ($x=0$) at $t=t_c$ and
the collision point moves to the larger-$|x|$ region.
(b) The behavior of collision of two 1-branes
by use of the different time $T=\sqrt{t^2-x^2}$.
In this frame two 1-branes collide simultaneously.
}
\label{collision_3D}
\end{center}
\end{figure}

For $\Lambda<0$,  the proper distance  for fixed $x$
 decreases as
$t$ increases from $t=0$, and it eventually vanishes at $t=t_c$.
Hence, one 1-brane approaches the other as time progresses,
causing the complete collision at $t=t_{\rm c}$.
If we fix the brane charges such that $M_1+M_2>0$,
the branes first collide at $x=0$, and
as the time evolves, 
the subsequent collisions
occur at the larger $|x|$
[see Fig. \ref{collision_3D}(a)].
This behavior, however, changes
by use of a different time coordinate.
For example, if we watch the collision
in the frame $T=\sqrt{t^2-x^2}$,
the collision occurs simultaneously
[see Fig. \ref{collision_3D}(b)].

On the other hand, when $\Lambda>0$,  the
proper distance with the fixed $x$ takes
the minimum value
\begin{eqnarray}
d_{\rm min}(x)=\frac{1}{2}L\left[-\Lambda\,x^2+
 \left(M_1+M_2\right)L\right]\,,
\end{eqnarray}
at $t=0$,
if  $|x|<x_{\rm max}$,
where $x_{\rm max}=\sqrt{(M_1+M_2)/\Lambda}$, and
the distance $d$
increases as $t$ increases.
For the region of $|x|>x_{\rm max}$, two branes are initially disconnected,
but they are connected at $t_c$ as the time evolves and the
distance also increases in time.
Then, for $t>0$, each brane gradually
separates the others as the time progresses.
This is similar to the case of $\Lambda>0$ for $D>4$.

\subsection{The dynamical 0-brane solution}
\label{sec:D0}

In this section,
we consider the case of $p=0$, i.e.,
a 2-form field strength in the action
(\ref{D:action:Eq}) in order to obtain a homogeneous
 cosmological solution.
Since the world volume for 0-brane has only the time coordinate, the
warp factor in the metric can depend on the time and transverse
space coordinates.

From Eqs. (\ref{D:h1:Eq}),
(\ref{D:Ricci-mn:Eq})
and (\ref{D:Ricci-ij:Eq}),
we find the basic equations in the present case
as
\Eqrsubl{D0:Einstein2:Eq}{
&&\pd_t^2h_0-\frac{4}{D-1}\Lambda=0,
   \label{D0:Ricci-mn:Eq}\\
&&R_{ij}(\Ysp)-\frac{2}{D-1}\Lambda u_{ij}=0,
   \label{D0:Ricci-ij:Eq}\\
&&
\lap_{\Ysp}h_1=0.
   \label{D0:h_1:Eq}
 }
Integrating Eq.~(\ref{D0:Ricci-mn:Eq}) , we find
\Eq{
h_0(t)=\frac{2\Lambda}{D-1} t^2+c_1t+c_2
\,,
   \label{D0:h0:Eq}
}
where $c_1$ and $c_2$ are integration constants.
Using a freedom of the time translation, we can shift
$c_1$. $c_2$ is included in $h_1$.
Hence, we can set $c_1=c_2=0$
without loss of generality, which we assume in what follows.

Equation~(\ref{D0:Ricci-ij:Eq}) means that the $\Ysp$ space 
is not Ricci flat due to the existence of a cosmological constant.
It is different from a class of time-dependent solutions with
multiple charged ``singular" solutions
coupled to a scalar field~\cite{Maki:1992tq}.
The harmonic function $h_1$ is obtained by solving
Eq. (\ref{D0:h_1:Eq}).
Although one can superpose the harmonic eigenfunctions
to find general solution of
$h_1$, those eigenfunctions should be solved on an
Einstein space, which may not be so trivial even if
$\Ysp$ is assumed to be a constant curvature space.
Without specifying the harmonic function $h_1$,
let us consider the dynamics of the metric in more detail.

Assuming $\Lambda>0$ and
introducing a new time coordinate, $\tau$ becomes
\Eq{
{\tau}=(D-2)
\, t^{1\over D-2}\,,
}
and we find the $D$-dimensional metric (\ref{D:metric:Eq}) as
\Eqr{
ds^2=
\left[1+\left(\frac{\tau}{\tau_0}\right)^{-2(D-2)}h_1
\right]^{-\frac{D-3}{D-2}}
\left[-d\tau^2+
\left\{1+\left(\frac{\tau}{\tau_0}\right)^{-2(D-2)}h_1\right\}
\left(\frac{\tau}{\tau_0}\right)^2u_{ij}dy^idy^j\right],
 \label{D0:s-metric:Eq}
 }
where $\tau_0=(D-2)\sqrt{(D-1)/(2\Lambda)}$.
When we set $h_1=0$, the spacetime is
an isotropic and homogeneous universe,
whose scale factor is proportional to $\tau$.
The $D$-dimensional spacetime becomes inhomogeneous
unless $h_1=0$. Thus, in the limit when the
terms with $h_1$ are negligible, which is realized in the limit
$\tau\rightarrow\infty$ for $D>2$,
we find a $D$-dimensional Milne universe,
which is guaranteed by a scalar field with
the exponential potential as we discussed in \ref{D1}.
It is interesting to note that the power exponent of the
 scale factor is always larger than that in
the matter dominated era or in the radiation dominated era.

In the case of $\Lambda<0$,
time $t$ is bounded by
$|t|<t_s$, where
\begin{eqnarray}
t_s(y)=\sqrt{(D-1)h_1(y)\over 2|\Lambda|}
\end{eqnarray}
 is the time when a singularity appears at $y$.
Thus if we assume that $h_1(y)>0$,
 the spacetime has initially no singularity,
but the Universe collapses to a big crunch,
which happens  at the different time $t_s(y)$
at each spatial point $y^i$.
\section{Conclusion}
  \label{sec:Conclusion}

We have derived the dynamical $p$-brane
solutions with a cosmological constant
and discussed their applications to brane collision and  cosmology.
These solutions were obtained by adding a cosmological
constant $\Lambda$ in the $D$-dimensional $p$-brane action
\cite{Gibbons:2005rt}.
The basic idea was to consider field configurations in higher
dimensions that are obtained by replacing the constant
in supersymmetric $p$-brane solutions with warped compactifications,
by a field on the world volume spacetime of the $p$-brane.
The resulting $D$-dimensional metric and the $(p+2)$-form
field strength depend not only on the quadratic function of the
time but also on that of spatial coordinates 
of the world volume of the $p$-brane.
We could never neglect the coordinates of world volume
if we add a cosmological constant. Thus, we find that
the form of $D$-dimensional metric is similar to that
of the $p$-brane with a trivial or vanishing dilaton.
The difference from the $p$-brane
metric is that the $(D-p-1)$-dimensional transverse spacetime
$\Ysp$ is an Einstein space and is not
in general Ricci flat except for the case of the 1-branes.
Moreover, the solution tells us that the function $h$ depends on all
the world volume coordinates of the $p$-brane. Hence, the contribution of
the field strength except for the 2-form leads to an inhomogeneous universe.

We have discussed the dynamics of 1-branes in the $D$-dimensional theory.
Since the curvature for the solution is singular at the places where $h=0$,
we should consider a domain where the metric is regular.
In the asymptotic far-brane region,
for a positive cosmological constant
the 1-brane spacetime with $D>4$
approaches the $D$-dimensional Milne universe,
while for a negative cosmological constant it does
a conformally flat and static inhomogenenous spacetime.
On the other hand, in regions close to the branes, for concreteness,
we have considered the case of two branes in detail.
For a positive cosmological constant, one 1-brane is approaching the other
as the time evolves for $t<0$ but separates the other for $t>0$.
In the case of $D>4$,
for a negative cosmological constant, we have found that
all of the domain between the branes are initially connected, but
some region (near small $x$) shrinks as the time increases, and
eventually the topology of the spacetime changes such that parts of
the branes are separated by a singular region surrounding each 1-brane.
Thus, in the case of $D>4$ 1-branes never collide.
On the other hand,
the case of $D=3$,
for a negative cosmological constant and $t>0$,
could provide an example of colliding branes.
We found that
the collision time depends on both brane charges and the place in the world volume.
As we illustrated in Fig. 7,
the collision process was observer-dependent.

Finally,
we have also used the 0-brane solutions with a 2-form field strength
and a cosmological constant to study cosmology.
In the case of $D=4$, the scale factor of our four-dimensional
spacetime is a linear function of the cosmic time
which is the same evolution as the Milne universe.
Without a cosmological constant,
we get the cosmic evolution in the radiation dominated universe.
Thus, although adding a cosmological constant helps
to obtain the expansion law which is closer to the realistic one,
it turned out that it was not still enough.

Although one might think that
the examples considered here may not
provide realistic cosmological models,
this is inevitable in such
fundamental theories like supergravities unless we also introduce
more matter fields and others.
However,
the properties we have discovered would give a
clue to investigate
cosmological models in more realistic higher-dimensional
cosmological settings.

\section*{Acknowledgments}

K.U. would like to thank H. Kodama, M. Sasaki, and T. Okamura
for continuing encouragement.
The work of K.M. was partially supported
by the Grant-in-Aid for Scientific Research
Fund of the JSPS (Grant No.22540291)
and by the Waseda University Grants for Special Research Projects.
The work of N.O. was supported in part by the Grant-in-Aid for
Scientific Research Fund of the JSPS (C) No. 20540283, No. 21$\cdot$09225 and
(A) No. 22244030.
K.U. is supported by Grant-in-Aid for Young Scientists (B) of JSPS Research, under
Contract No. 20740147.



\end{document}